\documentclass[preprint2]{aastex}

\slugcomment{Revised version 2011/06/01}

\shorttitle{\textit{AKARI}  Brown Dwarfs}
\shortauthors{Tsuji, Yamamura, and Sorahana}

\begin{document}


\title{\textit{AKARI} OBSERVATIONS OF BROWN DWARFS. II. CO$_2$ as Probe
of Carbon and Oxygen Abundances in Brown Dwarfs  \\ 
}



\author{Takashi Tsuji}
\affil{Institute of Astronomy, School of Science, The University of Tokyo, \\
2-21-1 Osawa, Mitaka, Tokyo 181-0015, Japan}
\email{ttsuji@ioa.s.u-tokyo.ac.jp}

\and

\author{Issei Yamamura and Satoko Sorahana\altaffilmark{1}}
\affil{Institute of Space and Astronautical Science (ISAS), JAXA,\\
Yoshino-dai 3-1-1, Chuo-ku, Sagamihara, Kanagawa 252-5210, Japan}
\email{yamamura@ir.isas.jaxa.jp, sorahana@ir.isas.jaxa.jp}

\altaffiltext{1}{Also at Department of Astronomy, School of Science, The University of Tokyo, 
Hongo 7-3-1, Bunkyo-ku, Tokyo 113-0033, Japan}




\begin{abstract}
Recent observations with  the infrared astronomical satellite \textit{AKARI} 
 have shown that the CO$_2$ bands at 4.2\,$\mu$m in three brown dwarfs 
are much stronger than expected from the unified cloudy model (UCM) 
based on  recent solar C \& O abundances. This result has been a puzzle,
but we now find that this is simply an abundance effect:  We show  that 
these strong CO$_2$ bands can be explained  with the UCMs based on the
classical C \& O abundances (log\,$A_{\rm C}$ and log\,$A_{\rm O}$),
which are about 0.2\,dex larger compared to the recent values. 
Since three other brown dwarfs could  be well interpreted with 
the recent solar C \& O abundances, we  require 
at least two model sequences based on the different chemical 
compositions to interpret all the \textit{AKARI} spectra. 
The reason for this is that the CO$_2$ band is especially 
sensitive to C \& O abundances, since the CO$_2$ abundance  
depends approximately on $A_{\rm C}A_{\rm O}^{2}$ --- the cube of
C \& O abundances. For this reason, even low 
resolution spectra of very cool dwarfs, especially of CO$_2$,
cannot be understood unless a model with proper abundances is applied.
For the same reason, CO$_2$ is an excellent indicator of 
C \& O abundances, and we can now estimate C \& O abundances of 
brown dwarfs: Three out of six brown dwarfs observed with \textit{AKARI}
 should have high C \& O abundances similar to
the classical solar values (e.g. log\,$A_{\rm C} = 8.60$
and log\,$A_{\rm O} = 8.92$), but the other three  may have low
C \& O abundances similar to the recent solar values
(e.g. log\,$A_{\rm C} = 8.39$ and log\,$A_{\rm O} = 8.69$). This result 
implies that three out of six brown dwarfs  are highly metal rich 
relative to the Sun if the recent solar C \& O abundances are correct. 
       
\end{abstract}



\section{INTRODUCTION}

In our recent work, the CO$_2$ molecule was identified for the first
time in the spectra  of brown dwarfs observed
with the infrared astronomical satellite 
\textit{AKARI} \citep[][hereafter referred to as Paper~I]{Yamamura10}.
We tried to interpret the observed behavior of the CO$_2$ band
with the use of the  model photosphere of brown dwarfs,
referred to as the unified cloudy model \citep[UCM;][]{Tsuji02, Tsuji05}. 
In modeling the photospheres of brown dwarfs, one problem
is how to consider the chemical composition, since no
direct abundance analysis is known for brown dwarfs. 
We thought it reasonable to assume a typical composition
for the disk  stellar population  such as the Sun. However, the solar 
composition itself experienced drastic changes in the past 
decades, and the true chemical composition of the Sun is still
by no means well established. Nevertheless,  we thought it appropriate 
to use the latest version of the solar abundances as possible 
proxies for the chemical abundances in the brown dwarfs. 
 
In our earlier version of UCMs \citep{Tsuji02}, we 
assumed the solar abundances largely based on the classical
LTE analysis using  one dimensional (1D) hydrostatic model 
photospheres and, in particular, C \& O abundances
were  log\,$A_{\rm C} = 8.60$ and log\,$A_{\rm O} = 8.92$ on the
scale of log\,$A_{\rm H} = 12.0$ \citep[e.g.][]{Anders89, Grevesse91}.
At about the same time as we were  computing our first version
of UCMs, a new result for the solar C \& O abundances 
based on three dimensional (3D)  time-dependent
 hydrodynamical model of the solar photosphere was published 
\citep{AllendePrieto02}. Since the classical 1D model
may be too simplified for the real solar photosphere,
this new approach seemed to be a useful contribution to the
solar abundance analysis.  The current version of UCMs 
\citep{Tsuji05} is thus based on this new result
(log\,$A_{\rm C} = 8.39$ and log\,$A_{\rm O} = 8.69$)  by 
\citet{AllendePrieto02} as noted elsewhere \citep{Tsuji04}.
 The new C \& O abundances are about 0.2\,dex smaller
 as compared to the classical values referred to above.

 We applied our current version of UCMs to 
the  brown dwarfs  observed with \textit{AKARI} in
Paper~I, and we could explain about half of our sample of spectra 
almost perfectly.  For the other half of our targets, we could explain 
the overall SEDs by this version of UCMs, but we could not 
explain their strong CO$_2$ bands. 
One explanation is that this may be due to an unknown process
related to CO$_2$, since anomalously strong CO band depths have
also been explained by a special process now known as vertical 
mixing \citep[e.g.]
[]{Noll97, Oppenheimer98, Griffith99, Saumon00, Leggett07b}.

However, we happened to try  our old version of UCMs based on the 
classical  C \& O abundances and  found that the
CO$_2$ band appeared to be much stronger in the spectra based on the 
old models than on the present models. At the first
glance, this is rather surprising  because C \& O abundances in 
the old models are only about 0.2\,dex larger than those in the
present models. However, we realized immediately that the CO$_2$ 
abundance is extremely  sensitive  to both C \& O abundances because 
the CO$_2$ abundance depends on the cube of C \& O abundances
($A_{\rm CO_2} \propto A_{\rm C}A_{\rm O}^{2}$). We recall that
the strong dependence of the CO$_2$ abundance on metallicity, [Fe/H]
\footnote{The differential iron abundance of a star relative to the Sun 
and defined by [Fe/H]= log${ (A_{\rm Fe}/A_{\rm H})_{*}}$ - 
log${ (A_{\rm Fe}/A_{\rm H})_{\odot}}$.}, was previously known 
by a detailed thermochemical analysis of the C, N, and O bearing
gaseous molecules \citep{Lodders02}.

The above result demonstrates that at least two different
series of model photospheres are needed for the analysis of
the CO$_2$ band observed with  \textit{AKARI}. For this purpose, 
we reconsider our old version of UCMs based on the classical
C \& O  abundances to represent a case of rather high C \& O
abundances. Our current version of UCMs based on the new C \& O 
abundances  will serve as representing a case of the reduced C \& O  
abundances  compared to the old version of the UCMs.
An important implication of this result is that the metallicity
(C \& O abundances) in brown dwarfs should have  a variety of values.
  
The interpretation and analysis of the spectra of cool dwarfs already 
have a rather long history \citep[e.g.][]{Kirkpatrick05, Burgasser06a, 
Leggett07a, Cushing08, Stephens09, Yamamura10}, and
the effect of metallicity has been discussed by some authors.
For example, \citet{Burgasser06b} have measured the strengths of the major
H$_2$O and CH$_4$ bands in the 1.0--2.5\,$\mu$m region in a large sample 
of T dwarfs, and found that  the  resultant spectral indices 
plotted against  spectral type  revealed considerable scatter.
Several reasons for this result including the effects of dust,
 gravity, and metallicity  have been considered, but it appeared
difficult to separate the effect of metallicity  from the remaining 
paremeters. \citet{Leggett09} have shown that the effect of metallicity 
on the SEDs of T dwarfs should be  significant, but noted that other 
parameters such as gravity can affect the SEDs similarly.
This result again showed the difficulty in determining metallicity
uniquely from SEDs.
Also, the so-called ``blue'' L dwarfs classified as L subdwarfs
have been interpreted to have low metallicity with [Fe/H] 
from $-1.5$ to $-1.0$ \citep[e.g.][]{Burgasser09}, but those 
L dwarfs with unusually blue near-infrared colors can also be 
explained by a patchy cloud model 
\citep{Folkes07, Marley10}.
  
The brief survey outlined above reveals that the problem of metallicity 
in brown dwarfs is still unresolved. In this paper, we will show  
clear evidence of metallicity
variations in brown dwarfs for the first time.
In fact,  the most important significance
of the discovery of CO$_2$ with \textit{AKARI} is that it
demonstrated the variations of the C \& O abundances by at least
50\% in brown dwarfs and that it  provided
a  means by which to estimate the C \& O abundances
in very cool dwarfs.

In  Paper I, we have analyzed the \textit{AKARI} spectra and  discussed  
the basic physical parameters of our objects in detail.  
There we have applied the conventional
method based on a direct comparison of the observed and predicted
spectra. In this paper, we examine the results of Paper~I by a more
detailed numerical method in Section~\ref{sec:fitcmp}, and we confirm 
that the physical
parameters determined in Paper~I mostly agree with those based on the
reduced-chi-square minimization method  within the estimated errors.
A problem, however, is that an adequate application of such a rigorous
numerical method
requires the input data of sufficient accuracy. Unfortunately, the
input data --- our present models of brown dwarfs --- are not
precise enough for this purpose, as discussed in Section~4.5 of Paper I.
Therefore the numerical method does not necessarily 
provide  the best answer and  
the traditional fitting method ``by eye'' can still be  useful for some cases. 
For these reasons, we use the physical parameters
determined in Paper~I and adopt the same approach as Paper~I,
eye-fitting, throughout this paper.

\begin{table*}
\caption{Carbon and Oxygen Abundances in UCMs }\label{tbl:tbl1}
\begin{center}
\begin{tabular}{cccl}
\noalign{\bigskip}
\hline
\noalign{\bigskip}
  series & log $A_{\rm C}$ $^{a}$  & log $A_{\rm O}$ $^{a}$ &  Note on
 chemical composition \\
\noalign{\bigskip}
\hline
\noalign{\bigskip}
UCM-a  &  8.60  &  8.92 & 1D solar abundances \citep[e.g.][]{Anders89}$^{b}$ \\ 
UCM-c &  8.39  &  8.69 & 3D solar abundances \citep[e.g.][]{AllendePrieto02}$^{c}$ \\
\noalign{\bigskip}
\hline
\end{tabular}
\end{center}
Notes.\\
  a) The logarithmic abundance on the scale of log $A_{\rm H}$ = 12.0. 
  b) Actually, we have applied a slightly updated version as summarized
    in Table~1 of \citet{Tsuji02}.
  c) A full listing of the abundances is given in 
http://www.mtk.ioa.s.u-tokyo.ac.jp/$\sim$ttsuji/export/ucm/tables/table1.dat.
 
\end{table*}

\begin{figure}[!ht]
  \begin{center}
   \resizebox{1.0\hsize}{!}{
       \includegraphics{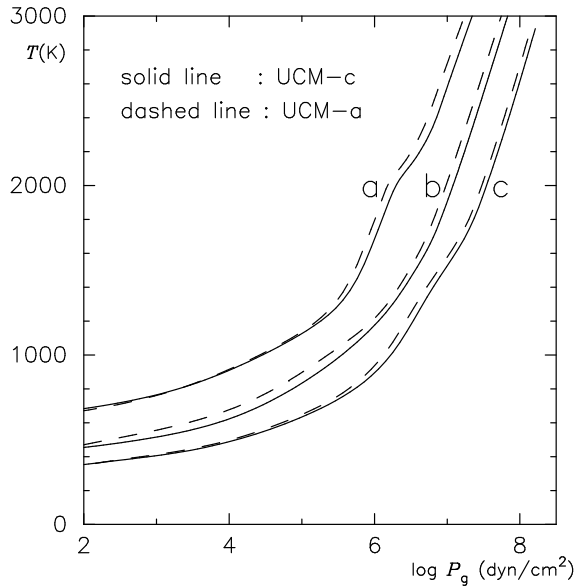}
   }
  \end{center}
\caption{The effects of C \& O abundances on the thermal 
structure of cool dwarfs. The solid and dashed lines are based on the 3D 
and 1D solar abundances (see Table~\ref{tbl:tbl1}), respectively.
a)  $T_\mathrm{eff}$ = 1500\,K, $T_{\rm cr}$ = 1700\,K, and $\log g = 4.5$.
b)  $T_\mathrm{eff}$ = 1200\,K, $T_{\rm cr}$ = 1900\,K, and $\log g = 4.5$.
c)  $T_\mathrm{eff}$ =  900\,K, $T_{\rm cr}$ = $T_{\rm cond}$,
 and $\log g = 4.5$.
}\label{fig:fig1}
\end{figure}

\begin{figure}[!ht]
  \begin{center}
   \resizebox{1.0\hsize}{!}{
       \includegraphics{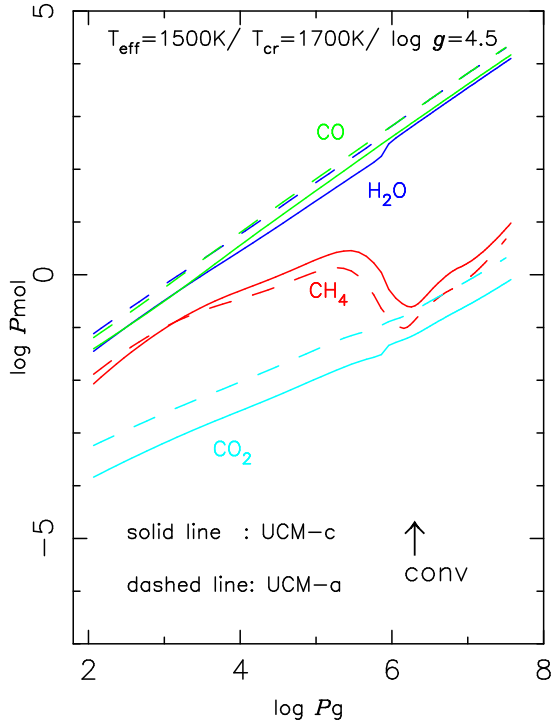}
   }
  \end{center}
\caption{The effects of C \& O abundances on the
 molecular abundances in cool dwarfs. 
The solid and dashed lines are based on the 3D 
and 1D solar abundances (see Table 1), respectively.
The physical parameters of the model are:
 $T_\mathrm{eff}$ = 1500\,K, $T_{\rm cr}$ = 1700\,K, and $\log g =
 4.5$. This case may apply to 2MASS J152322$+$3014 (solid line) and 
SDSS J083008$+$4828 (dashed line). The units of $P_{g}$ and 
$P_{\rm mol}$ are dyn\,cm$^{-2}$. The arrow indicates the onset of convection. 
}\label{fig:fig2}
\end{figure}

\begin{figure}[!ht]
  \begin{center}
   \resizebox{1.0\hsize}{!}{
       \includegraphics{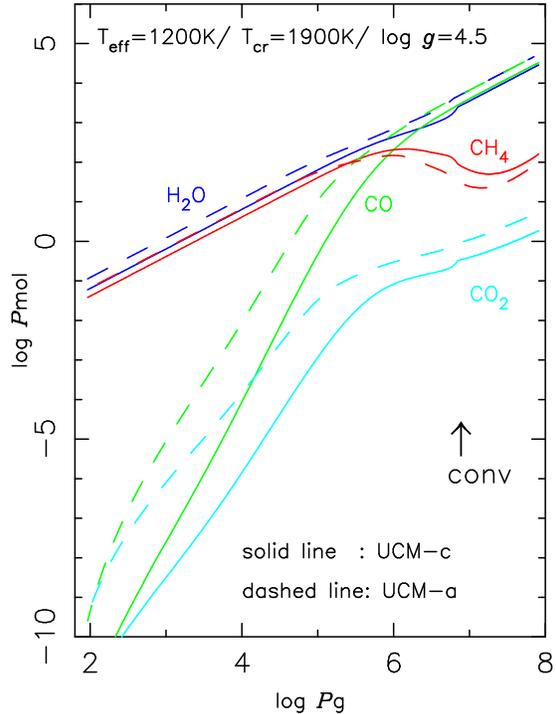}
   }
  \end{center}
\caption{The same as Figure~\ref{fig:fig2} but for the physical
 parameters of the model:
  $T_\mathrm{eff}$ = 1200\,K, $T_{\rm cr}$ = 1900\,K, and $\log g =
 4.5$. This case may apply to 2MASS J055919$-$1404 (dashed line).  
}\label{fig:fig3}
\end{figure}

\begin{figure*}[!ht]
  \begin{center}
   \resizebox{0.98\hsize}{!}{
       \includegraphics{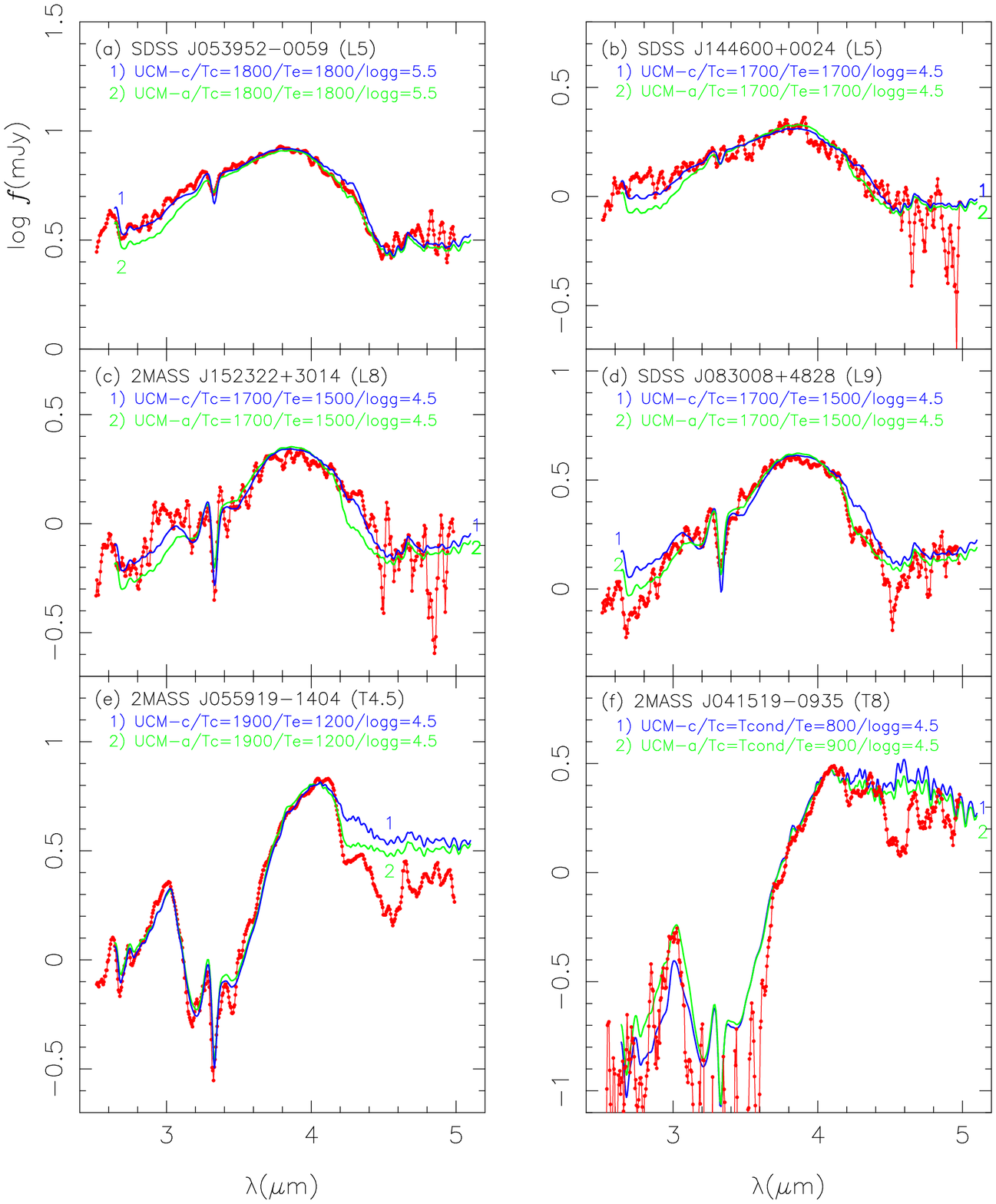}
   }
  \end{center}
\caption{Spectra observed with \textit{AKARI} are compared with the 
predicted spectra based on the models of the UCM-c (curve~1, blue) and UCM-a
(curve~2, green) series.
The UCM-a series is based on the classical solar C \& O abundances (1D
abundances, see Table~\ref{tbl:tbl1}) and the UCM-c series on the more 
recent solar C \& O abundances (3D abundances, 
see Table~\ref{tbl:tbl1}). Note that the brown dwarfs
shown in (a), (b), and (c) are relatively well fitted by the
models of the UCM-c series while those in (d), (e), and (f)
are better described by the models of the UCM-a series.
}\label{fig:six}
\end{figure*}

\section{ROLE OF THE CARBON AND OXYGEN ABUNDANCES IN THE UCM}
 Our two series of  UCMs  are referred to
as UCM-a and UCM-c, and  they only differ in C \& O abundances 
as summarized in Table~\ref{tbl:tbl1}. The UCM-a series is based on the 
classical C \& O abundances, which we refer to as 1D solar 
abundances for simplicity, and the UCM-c series on the new abundances, which   
we refer to as 3D solar abundances. 

We first examine the effects of C \& O abundances
on the thermal structure of the photosphere. For this purpose, 
the models of the UCM-c series are taken from our database\footnote{The 
numerical details of this version (UCM-c series) are available from 
http://www.mtk.ioa.s.u-tokyo.ac.jp/$\sim$ttsuji/export/ucmLM/ and
/ucm/.}. 
Since our code to compute UCMs has been modified to some extent  over 
the last 10 years, we recompute all the models of the UCM-a series  used 
in the present paper. Therefore
the models of the UCM-a and UCM-c series are 
now computed by exactly the same code, except for the Rosseland and 
Planck mean opacities,  
which  of course differ according to the chemical composition adopted.
  
We show a simple comparison of the photospheric structures of 
the UCM-a and UCM-c series
for the cases of $T_\mathrm{eff}$  = 900, 1200, and 1500\,K in 
Figure~\ref{fig:fig1}. Other parameters such as $T_\mathrm{cr}$ and 
log $g$ are chosen to be those actually found for our objects (see 
Table~\ref{tbl:fit}). Inspection of Figure~\ref{fig:fig1} reveals that 
the models of the UCM-a series shown by the dashed lines are generally 
warmer by up to about 100\,K as compared to the models of the 
UCM-c series shown by the solid lines. Since the major
opacity sources such as CO and H$_2$O 
are more abundant in the UCM-a than in the UCM-c series, the blanketing 
effect due to molecular bands should be more effective and hence 
the models of the UCM-a series are warmer than those of the UCM-c series.

Next, we examine the effects of C \& O abundances on the 
CO$_2$ and other molecular abundances. We present the 
abundances of H$_2$O, CO, CO$_2$, and CH$_4$  for the case of
$T_\mathrm{eff}$ = 1500\,K, $T_{\rm cr}$ = 1700\,K, and $\log g = 4.5$ 
for the models of the UCM-c ( applied to 2MASS J152322$+$3014 in 
Section~\ref{sec:spc}) and UCM-a (applied to SDSS J083008$+$4828) 
series in Figure~\ref{fig:fig2} as the solid 
and dashed lines, respectively. The increased C \& O 
abundances result in the increases of CO, CO$_2$, and H$_2$O abundances
as expected. The increase of the CO$_2$ abundance in the UCM-a series is 
quite significant for the reason noted before. 
On the contrary, the CH$_4$ abundance shows a decrease in the UCM-a
series and this unexpected result may be because the direct 
effect of the increased carbon abundance on the CH$_4$ abundance is 
superseded by the dissociation of  CH$_4$ due to the elevated  
temperatures in the model of the UCM-a series (Figure~\ref{fig:fig1}).

As another example, we show the case of $T_\mathrm{eff}$ = 1200\,K, 
$T_{\rm cr}$ = 1900\,K, and $\log g = 4.5$ in Figure~\ref{fig:fig3}. 
The results are again shown for the UCM-c and UCM-a series
with the solid and dashed  lines, respectively. In this case, the 
effects of the abundance changes are more pronounced, especially for CO$_2$.  
We will see in Section~\ref{sec:spc} that the case of the UCM-a series is
approximately realized in 2MASS J055919$-$1404 in which the CO$_2$ band 
appears to be very strong.

\begin{figure}[!ht]
  \begin{center}
   \resizebox{1.0\hsize}{!}{
       \includegraphics{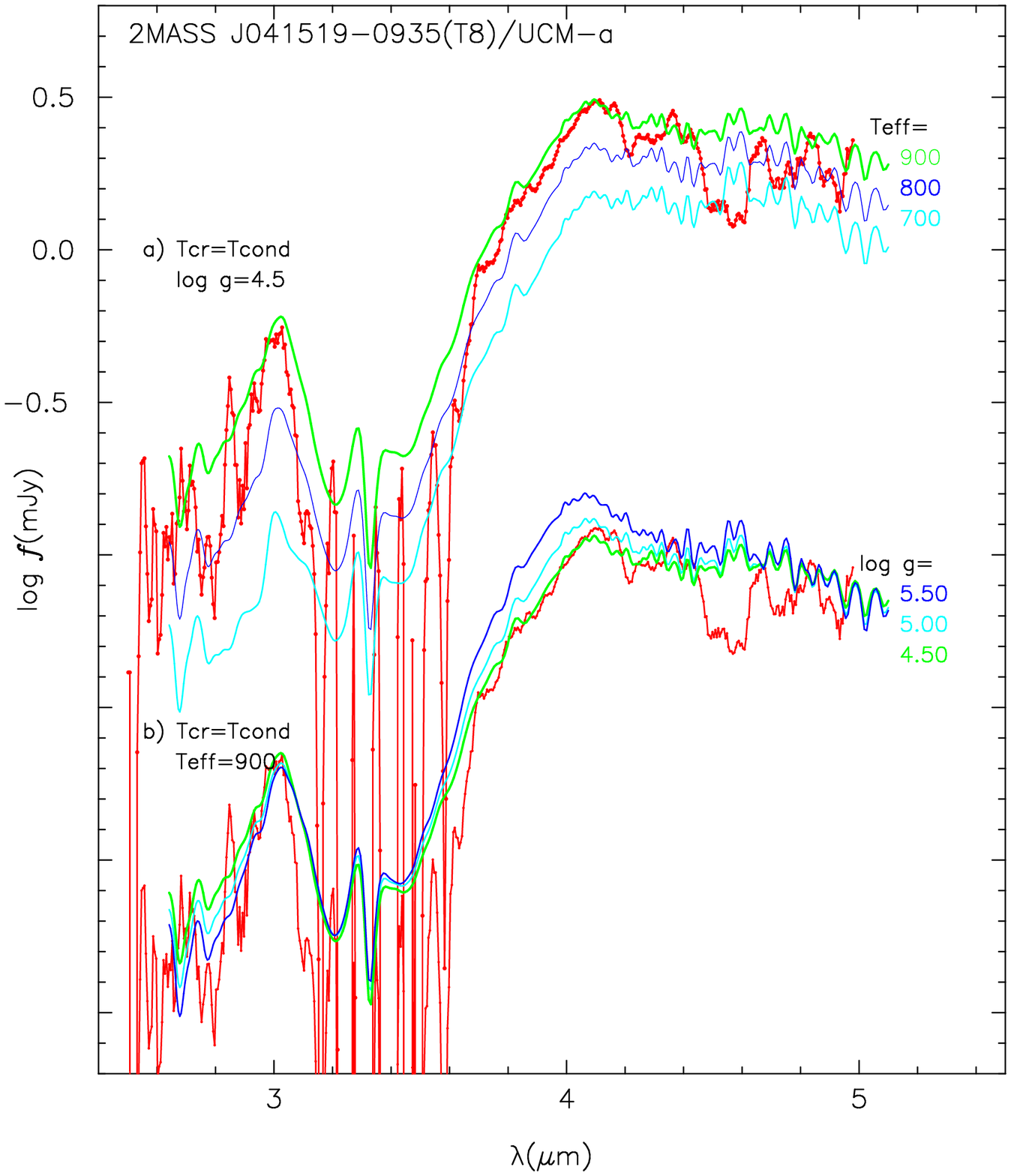}
   }
  \end{center}
\caption{Observed spectrum of 2MASS J041519$-$0935
is compared with the predicted spectra based on the models of
the UCM-a series. The best fit is obtained for
$T_\mathrm{eff}$ = 900\,K, $T_{\rm cr} = T_{\rm cond}$, 
and $\log g = 4.5$.
a) Effect of $T_{\rm eff}$ under the fixed values
of $T_{\rm cr} = T_{\rm cond}$ and log $g$ = 4.5.  
b) Effect of log $g$ under the fixed values
of $T_{\rm eff} = 900$\,K and  $T_{\rm cr} = T_{\rm cond}$.
}\label{fig:j0415}
\end{figure}

\section{EFFECTS OF THE CARBON AND OXYGEN ABUNDANCES ON
 THE SPECTRA OF BROWN DWARFS}\label{sec:spc}

We compare the observed spectra of six brown dwarfs with the
predicted ones based on the models of the UCM-a and UCM-c series
in Figure~\ref{fig:six} (a)--(f). The effect of C \& O 
abundances on the spectra of brown dwarfs  can be seen  most clearly in a
comparison of 2MASS J152322$+$3014 and  SDSS J083008$+$4828 shown in 
Figure~\ref{fig:six}(c) and Figure~\ref{fig:six}(d), respectively. These 
two brown dwarfs were found to have  nearly the same physical
parameters ($T_{\rm eff} = 1500$\,K, $T_{\rm cr} = 1700$\,K, and
log\,$g$ = 4.5)  but the spectra looked to be quite different
(Paper~I).  In particular, the CO$_2$ band at 4.2\,$\mu$m appeared to be
much stronger in SDSS J083008$+$4828 than in 2MASS J152322$+$3014. 
In 2MASS J152322$+$3014, the observed spectrum could be 
accounted for by the model of the UCM-c series (Paper I), as confirmed
by curve~1 in Figure~\ref{fig:six}(c):     
Especially, the regions of the H$_2$O 2.7\,$\mu$m  and the CO$_2$ 4.2\,$\mu$m 
bands as well as the $Q$-branch of CH$_4$ band appeared to be well
explained by the model of the UCM-c series. On the other hand, the
predicted spectrum based on the model of the UCM-a series shown 
by curve~2 can not explain those features. 

In SDSS J083008$+$4828 shown in Figure~\ref{fig:six}(d), the observed 
spectrum could not be accounted for by the model of
the UCM-c series (Paper~I), as confirmed by  curve~1.  On the other hand, 
the strong CO$_2$ band at 4.2\,$\mu$m can now be 
explained reasonably well by the model of the UCM-a series, as
 shown by curve~2 in Figure~\ref{fig:six}(d). Thus the large depression 
due to the CO$_2$ band turns out to be due to   
the high C \& O abundances in a LTE model.
It is to be noted that this result is due to the increase   
of both C \& O abundances. In fact, a change of the 
C abundance alone, for example, produced 
only minor change on the CO$_2$ band strength (Paper~I). The large 
depression over the 2.7\,$\mu$m region mostly due to  H$_2$O 
can also be better explained with the model of the UCM-a series.
On the contrary, the $Q$-branch of the CH$_4$ band
appears to be weaker with the model of the UCM-a than with that of 
the UCM-c series, consistent with
the decrease of the CH$_4$  in the UCM-a  compared to
the UCM-c series (Figure~\ref{fig:fig2}).  
Also, the observed CH$_4$ $Q$-branch indeed  agrees better 
with the predicted one based on the model of  UCM-a.
Thus, we conclude that the UCM-a series should be applied to 
SDSS J083008$+$4828 rather than the UCM-c series.

\begin{table*}
\caption{Basic Parameters from the Model Fittings by Using UCMs}\label{tbl:fit}
\begin{center}
\begin{tabular}{clcccccc}
\noalign{\bigskip}
\hline
\noalign{\bigskip}
 no. & object & UCM series$^{~a}$ & $T_\mathrm{eff}$\,(K) &
 $T_\mathrm{cr}$\,(K) & $\log g$ & 
$R/R_{J}^{~b}$ & $\Delta\,T_\mathrm{eff}$\,(K)$^{~c}$\\ 
\noalign{\bigskip}
\hline
\noalign{\bigskip}
1 & \object[SDSS J053951.99$-$005902.0]{SDSS  J053952$-$0059} & UCM-c & 1800 & 1800 & 5.5 & 0.804 & $-110$ \\ 
2 & \object[SDSS J144600.60$+$002452.0]{SDSS  J144600$+$0024} & UCM-c & 1700 & 1700 & 4.5 & 0.716 & $-108$ \\ 
3 & \object[2MASS  J15232263$+$3014562]{2MASS J152322$+$3014} & UCM-c & 1500 & 1700 & 4.5 & 0.684 & $-170$ \\
4 & \object[SDSS J083008.12$+$482847.4]{SDSS  J083008$+$4828} & UCM-a & 1500 & 1700 & 4.5 & 0.610 & $-173$ \\
5 & \object[2MASS  J05591914$-$1404488]{2MASS J055919$-$1404} & UCM-a & 1200 & 1900 & 4.5 & 1.122 & $-269$ \\
6 & \object[2MASS  J04151954$-$0935066]{2MASS J041519$-$0935} & UCM-a & ~900 & $T_\mathrm{cond}$ & 4.5 & 0.676 & $-136$ \\
\noalign{\bigskip}
\hline
\end{tabular}
\end{center}
Notes.\\
 a) The UCM series applied and indicates approximate C \& O abundances
    (see Table~\ref{tbl:tbl1}).
 b) Radius $R$ relative to the Jupiter's radius $R_{J}$.
 c) $\Delta\,T_\mathrm{eff} = T_\mathrm{eff}$\,(empirical values
by \citet{Vrba04}) 
 $-$  $T_\mathrm{eff}$\,(column 4 in this Table). 
\end{table*}   

In Figure~\ref{fig:six}(a), we compare the observed 
spectrum of SDSS J053952$-$0059 with the predicted ones based on the 
models of the UCM-a and UCM-c series. We already know that
this spectrum is well explained by a model of UCM-c series in Paper~I 
(curve~1).  On the other hand, the observed spectrum cannot be  
explained by a model of the UCM-a series of the same parameters 
(curve~2). 
We obtain more or less similar result for SDSS J144600$+$0024, namely the 
observed spectrum of this object can be well explained by
the model of the UCM-c (curve~1 in Figure~\ref{fig:six}(b)), but 
not with that of the UCM-a series (curve~2). 

In Figure~\ref{fig:six}(e), we examine the case of 2MASS J055919$-$1404 
in which the observed
CO$_2$ band is very strong. We could not explain the CO$_2$ and CO bands
in this object with our UCM-c series (Paper I), as confirmed  by
curve~1. But we can now explain the strong CO$_2$ band  approximately
with our model of the UCM-a series, as shown by 
curve~2. This result is fairly consistent with 
the very large increase of the CO$_2$ abundance for this model as noted 
in Figure~\ref{fig:fig3}.  The fit of curve~2 based on the UCM-a series 
can in principle be improved further 
by a fine tuning of C \& O abundances. 
But we defer such a detailed abundance analysis to 
future works and we only note here that the strength of the CO$_2$ band
is adjustable by changing C \& O abundances. 

Lastly, we examine the case of 2MASS J041519$-$0935 in 
Figure~\ref{fig:six}(f). This is a case in which the fitting parameters 
had to be changed from the result of Paper~I:  
We previously found that  $T_{\rm eff} = 800$\,K based on the model 
 of the UCM-c series.
However, the H$_2$O band at 2.7\,$\mu$m appears to be too strong if
we apply the same parameters to the model of the UCM-a series.
As shown in Figure~\ref{fig:j0415}(a), the observed 
spectrum can be  explained by the model
of a higher effective temperature of  $T_{\rm eff} = 900$\,K 
with the model of the UCM-a series. Also, we find that the case of 
log\,$g = 4.5$ provides the best fit as  shown in 
Figure~\ref{fig:j0415}(b). Thus, we conclude that ($T_{\rm eff}$, $T_{\rm cr}$,
log\,$g$) = (900, $T_{\rm cond}$, 4.5) for 2MASS J041519$-$0935 
for the models of the UCM-a series. 
We show the  best possible predicted spectrum based
on the model of the UCM-c series with $T_{\rm eff} = 800$\,K
and that based on the model of the UCM-a series with $T_{\rm eff} = 900$\,K 
by curve~1 and curve~2, respectively,  in Figure~\ref{fig:six}(f). We
see that  the model of the UCM-a series appears to match  better with
the observed spectrum  than the model of the UCM-c series.

Finally, we summarize the basic parameters of the six brown dwarfs
in Table~\ref{tbl:fit}. The major change from Table~4 of Paper~I is to have 
introduced abundance classes in column 3 indicated by the UCM series  
applied : UCM-a and UCM-c means that C \& O abundances should be close
to those of the 1D  and 3D solar  abundances, respectively. 
The physical parameters are re-examined with the models of the UCM-a
series  in the same manner  as in Paper~I.
A major change in physical parameter is  $T_\mathrm{eff}$ 
for 2MASS J041519$-$0935, which  is changed from 800\,K to 
900\,K as shown in Figure~\ref{fig:j0415}(a). As for other five objects, 
the physical
parameters for the models of the UCM-a series remain the same as for the 
UCM-c series. We confirm that the overall
SEDs based on the models of the UCM-a and UCM-c series of the same physical
parameters agree  well (see Figure~\ref{fig:six}) even though some
local features due to the CO$_2$ and H$_2$O bands differ somewhat. 
Thus, it is natural that the physical parameters mainly derived from the 
fits of the overall SEDs remain the same for UCM-a and UCM-c series.  
The $R/R_J$ values for SDSS J083008$+$4828, 2MASS J055919$-$1404, 
and 2MASS J041519$-$0935 are  changed  slightly, 
reflecting the changes of models from the UCM-c to the UCM-a series.

\section{DISCUSSION}

\subsection{Validity of our ``best fit'' models}\label{sec:fitcmp}
Fitting of the model spectra to the observed spectra is
carried out by ``eyes'' throughout in this paper as well as Paper~I.
The reader might wonder whether the eye-fitting can find reliable
``best'' models for various objects. In this subsection
we assess our eye-fitting by comparing with 
the numerical fitting results.

We evaluate the goodness of  fit by  the reduced-chi-square
(hereafter $\mathcal{R}$) defined as,
\begin{equation}
\mathcal{R} = \sum_{i=1}^{N} \left( \frac{f_{i} - C F_{i}}{\sigma_{i}}\right)^2 / \left(N-m\right),
\end{equation}
where $f_i$ and $F_i$ are fluxes of the observed and model spectra 
at $i$-th wavelength grid, respectively. The uncertainty of the 
observed flux is indicated as $\sigma_i$, and $m$ is the degree of freedom.
$C$ is the scaling factor that minimizes $\mathcal{R}$ and is given by
\begin{equation}
C = \frac{\sum f_{i} F_{i}/\sigma_{i}^2}{\sum {F_{i}}^2/{\sigma_{i}}^2}.
\end{equation}
These definitions are in principle equivalent to 
``Goodness-of-fit'' statistics $G$ by \citet{Cushing08}  for 
the equal weight case.

From our experience in Paper~I we know that the current UCM cannot 
fit the observations beyond 4\,$\mu$m at least in some objects. 
Therefore we limit the wavelength range
for calculating $\mathcal{R}$ to 2.64--4.15\,$\mu$m 
(cf. Model spectra are available from 2.64\,$\mu$m 
and CO$_2$ band starts from 4.17\,$\mu$m).

\begin{table}
\caption{The Best Three Models  According to the Numerical 
Fitting.}\label{tbl:fitcmp}
\begin{center}
\begin{tabular}{rccccc}
\hline
\hline 
No. & $T_\mathrm{eff}$ & $\log g$ & $T_\mathrm{cr}$ & $C$ & $\mathcal{R}$ \\
    & (K) & & (K) & ($\times 10^6$) \\
\hline
\hline
\multicolumn{6}{c}{SDSS J053951$-$0059 (L5)} \\
\hline 
{\bf 1} & {\bf 1800} & {\bf 5.5} & {\bf 1800} & {\bf 6.32} & {\bf 1.127}\\
  2 & 1900 & 5.5 & 1800 & 5.82 & 1.206 \\
  3 & 1900 & 5.0 & 1800 & 6.37 & 1.288 \\

\hline 
\hline 
\multicolumn{6}{c}{SDSS  J144600$+$0024 (L5)} \\
\hline 
  1 & 2000 & 4.5 & 1700 & 1.53 & 0.496 \\
  2 & 1900 & 4.5 & 1700 & 1.61 & 0.513 \\
  3 & 1800 & 4.5 & 1700 & 1.69 & 0.553 \\
{\bf 10} & {\bf 1700} & {\bf 4.5} & {\bf 1700} & {\bf 1.79} & {\bf 0.695} \\

\hline 
\hline 
\multicolumn{6}{c}{2MASS J152322$+$3014 (L8)} \\
\hline
  1 & 1600 & 5.5 & 1700 & 1.89 & 0.675 \\
  2 & 1600 & 5.0 & 1700 & 1.91 & 0.733 \\
  3 & 1700 & 5.5 & 1800 & 1.69 & 0.740 \\
{\bf  4} & {\bf 1500} & {\bf 4.5} & {\bf 1700} & {\bf 2.45} & {\bf 0.793} \\

\hline 
\hline 
\multicolumn{6}{c}{SDSS  J083008$+$4828 (L9)} \\
\hline 
  1 & 1600 & 4.5 & 1800 & 3.78 & 0.679 \\
  2 & 1700 & 4.5 & 1800 & 3.52 & 0.711 \\
  3 & 1800 & 5.0 & 1900 & 3.44 & 0.746 \\
{\bf  9} & {\bf 1500} & {\bf 4.5} & {\bf 1700} & {\bf 4.54} & {\bf 0.841} \\

\hline 
\hline 
\multicolumn{6}{c}{2MASS J055919$-$1404 (T4.5)} \\
\hline 
  1 & 1200 & 4.5 & $T_\mathrm{cond}$ & 21.8 & 0.389 \\
{\bf  2} & {\bf 1200} & {\bf 4.5} & {\bf 1900} & {\bf 20.3} & {\bf 0.418} \\
  3 & 1100 & 4.5 & 1700 & 25.3 & 0.482 \\

\hline
\hline 
\multicolumn{6}{c}{2MASS J041519$-$0935 (T8)} \\
\hline 
{\bf  1} & {\bf  800} & {\bf 4.5} & {\bf $T_\mathrm{cond}$} & {\bf 29.1} & {\bf 0.170} \\
  2 &  900 & 4.5 & $T_\mathrm{cond}$ & 20.6 & 0.173 \\
  3 &  900 & 5.0 & $T_\mathrm{cond}$ & 17.2 & 0.195 \\
\hline 

\end{tabular}
\end{center}
Notes.\\
Models of the UCM-c series are adopted throughout as in Paper I.
The eye-fitting results quoted from Paper~I are indicated  in bold-face.
\end{table}

In Table~\ref{tbl:fitcmp} we list the three best models
based on the $\mathcal{R}$ value and the model quoted in
Paper~I (by eye-fitting) for each object in our sample.
The eye-fitting results are indicated in bold-face.
The models selected by the eye-fitting achieve the minimum
$\mathcal{R}$ for SDSS J053951$-$0059 (L5) and 
2MASS J041519$-$0935 (T8; for this particular object we search 
for the best model among those of $T_\mathrm{cr} = T_\mathrm{cond}$
for the reason outlined in Section~4.3.6 of Paper~I)
and the second minimum $\mathcal{R}$ for 2MASS J055919$-$1404 (T4.5).
The difference of $\mathcal{R}$ between the first and second model
for the last case is tiny, and the model parameters are within 
the uncertainty we stated in Paper~I ($\pm 100$ K for $T_\mathrm{eff}$ 
and $T_\mathrm{cr}$, and $\pm0.5$ dex for $\log g$). 

For two late-L objects, 2MASS J152322$+$3014 (L8) and 
SDSS J083008$+$4828 (L9), the differences in the model parameters 
are mostly within the uncertainty of eye-fitting, although the models used in 
Paper~I are not included in the numerical best three for  
these dwarfs. Altough $\log g$ of 2MASS J152322$+$3014 differs 
by 1.0 dex, the eye-selected model is in the 4th position 
in the list, and we consider that it is still in the accepted range.

A significant difference between the two methods is found in the 
L5 dwarf SDSS J144600$+$0024. The numerical fitting suggests
$T_\mathrm{eff} = 2000$ K as the best, which is 300 K higher than the one 
selected by the eye-fitting. The second and third are of $T_\mathrm{eff}$
= 1900 and 1800 K. $\log g$ and $T_\mathrm{cr}$ are the same
in all fours models. In fact we regarded such high 
$T_\mathrm{eff}$ values to be unrealistic
for an L5 dwarf, and did not consider them in the eye-fitting.
The empirical $T_\mathrm{eff}$ derived by \citet{Vrba04} of this object
is even low as 1592 K.
A key feature is the CH$_4$ 3.3\,$\mu$m band, which 
appears only in the $T_\mathrm{eff} = 1700$ K model. 
The observed spectrum of this  source is rather noisy and the 
detection of this band is marginal.
If the tiny dip seen near 3.3\,$\mu$m in the observed spectrum is 
actually the CH$_4$ band, the eye-fitting results, even if it is not 
perfect, are justified.

The nature of mid-L to early-T type
dwarfs are still under debate and their effective temperatures 
might actually spread to higher values. Incomplete atmosphere 
modeling is another possible reason. As we discuss in Paper~I, 
current atmosphere models for brown dwarfs are still  
exploratory and the UCM is one of such models. 
There are many physical and chemical processes not yet understood in 
brown dwarfs. These problems shall be attacked and eventually
incorporated into  future model atmospheres, but
it is beyond the scope of the current paper.
In addition, some of our \textit{AKARI} spectra have relatively low S/N.
Under such circumstances, numerical fitting may not
always return a unique and physically reasonable solution. 
On the other hand the eye-fitting would give weight to some key 
features and consider balance over the wavelength range.
Our goal in this paper is to highlight the effects of chemical 
abundance  in the brown dwarf atmosphere. 
The comparisons described above well demonstrate 
that the eye-fitting is, even if it is not perfect, 
useful to find reasonable models for our purpose.
Therefore, we apply the model parameters based on our eye-fitting 
for the six objects including J144600$+$0024 in the analysis 
throughout this paper.

It is noted that the differences between the models appear much 
more prominently  over the shorter wavelength  range 
especially in $J$-band, even if the spectra in the \textit{AKARI} 
wavelength range  are similar to each other.
Consideration of near-infrared data such as 2MASS photometry or
ground-based spectroscopy will enable  constraining the model
parameters better (Sorahana et al. in preparation).
Such improvements in  spectral range will also help us to
evaluate the goodness of the fit in the wavelengths beyond 4\,$\mu$m.

\subsection{C \& O Abundances in Brown Dwarfs}

The very strong CO$_2$ feature observed with \textit{AKARI} in some
brown dwarfs has remained a puzzle (Paper~I), but we find that
this is simply due to the effect of C \& O abundances. 
Generally, a small change in the chemical composition does not 
have a large effect on the predicted spectra at low resolution
nor on the thermal structure of the photosphere in hotter stars. 
In fact, this is the reason why one dimensional spectral classification 
(e.g. Harvard system) is possible for such stars.  But in the case of cool 
stars, a small change of the chemical composition is amplified
in molecular abundances. A drastic example is the
spectral branching of cool giant stars into M, S, and C types 
according to whether  the C/O ratio is smaller or larger than unity.
In cool dwarfs, the change  of C \& O abundances also produces 
significant effect  on the  strengths of molecular bands as well as on the
photospheric structures because of the large molecular opacities.  

The results of Section~\ref{sec:spc} reveal that  half of the brown 
dwarf spectra observed with 
\textit{AKARI}  (i.e. SDSS J053952$-$0059, SDSS J144600$+$0024, and 
2MASS J152322$+$3014 ) can be fitted by the predicted spectra based on
the models of the UCM-c series  
(Paper~I). Although the fits are by no means perfect, the fits with the
predicted spectra based on the UCM-c series are better than those 
based on the UMC-a series. For this reason, C \& O abundances in these 
three brown dwarfs should be closer to the recent 3D solar abundances 
rather than to the classical 1D solar abundances. On the other hand,  
the remaining half (i.e. SDSS J083008$+$4828, 2MASS
J055919$-$1404, and 2MASS J041519$-$0935) of our sample 
can be reasonably accounted for by the models of the UCM-a series.
Therefore, C \& O abundances in these three objects should be 
closer to the classical 1D solar abundances rather than to the recent 
3D solar abundances. 

Since [Fe/H] of the main sequence stars in the Galactic disk
covers the range from $-0.8$ to +0.2 \citep[e.g.][]{Edvardsson93},
the same metallicity distribution may apply to brown dwarfs.
 It is certainly only by chance that the brown dwarfs we
have observed are divided into two groups by C \& O abundances.
Our sample is too small to investigate the metallicity distribution
in brown dwarfs, and we hope that this problem can be pursued
further with a larger sample.

The problem of the solar C \& O abundances  is still under intensive 
discussion  \citep[e.g.][]{Ayres06, Caffau08, Asplund09}.
Although  our problem here is not the solar composition,
it is of some interest to know which of the proposed
solar composition  results more realistic for the Sun.
If the recent 3D result is more realistic for the Sun, three of our
sample may have about  solar composition and the 
remaining three may be about 0.2\,dex more metal rich.
This means that the proportion of  metal rich
objects with the highest [Fe/H] of about +0.2 
is quite high in our present sample of brown dwarfs.

\subsection{Effects of C \& O abundances on the 0.9--2.5\,$\mu$m spectra}
\begin{figure}[!ht]
  \begin{center}
   \resizebox{1.0\hsize}{!}{
       \includegraphics{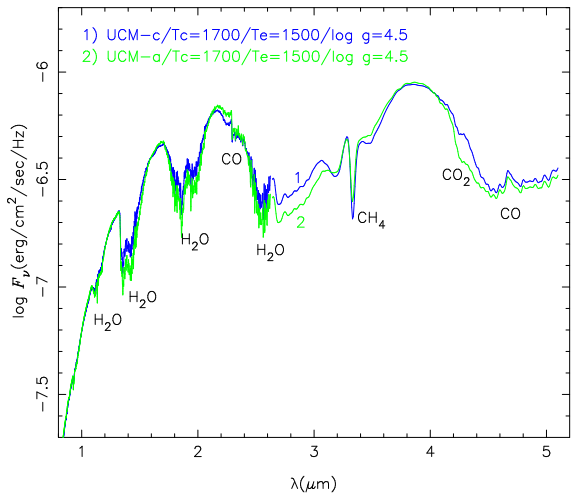}
   }
  \end{center}
\caption{ Comparison of the predicted spectra based on the models of
the UCM-a and UCM-c series for the case of $T_{\rm cr}$ = 1700\,K,
$T_\mathrm{eff}$ = 1500\,K,  and $\log g = 4.5$. The 0.9--2.5\,$\mu$m 
spectrum is convolved with the slit function of FWHM = 500\,km s$^{-1}$ 
(a typical resolution of observed spectra in this region), and the 
2.5--5.0\,$\mu$m spectrum with that of FWHM = 3000\,km s$^{-1}$ 
(the resolution of \textit{AKARI} spectra).
}\label{fig:nir}
\end{figure}

We have shown that  C \& O abundances have significant effects
on the 2.5--5.0\,$\mu$m spectra of brown dwarfs, and we now
examine their effect on the 0.9--2.5\,$\mu$m spectra. 
As  examples, we compare the predicted 0.9--2.5\,$\mu$m spectra of 
the models of UCM-a and UCM-c series for the case of $T_{\rm cr} = 1700$\,K, 
$T_{\rm eff} = 1500$\,K, and log\,$g$ = 4.5, together
with those for the 2.5--5.0\,$\mu$m spectra in Figure~\ref{fig:nir}.
The major difference between  
the UCM-a and UCM-c series is
that H$_2$O bands at 1.1, 1.4, 1.9, and 2.7\,$\mu$m are all
stronger in the UCM-a (curve~2 in Figure~\ref{fig:nir}) than in 
the UCM-c series (curve~1), and this is due to a direct effect of the 
increased oxygen abundance (see Figure~\ref{fig:fig2}).  

Thus the H$_2$O band strengths depend sensitively on oxygen abundance,
and we may hope to determine oxygen abundance from the H$_2$O bands.
However, we must remember that the H$_2$O band strengths
also depend on other parameters such as $T_{\rm cr}$, $T_{\rm eff}$, 
and log\,$g$, and we should encounter the same difficulty due to
a degeneracy of the parameters as noted before
by other authors \citep[e.g.][]{Burgasser06b, Leggett09}.
This fact reconfirms  the unique role of CO$_2$ as a
metallicity (C \& O abundances) indicator in brown dwarfs.

\section{CONCLUDING REMARKS}

Thanks to the \textit{AKARI} spectra, we are for the first time able to
demonstrate that the metallicity, more specifically C \& O
abundances, are important parameters to understand  brown
dwarf atmospheres. Until now, we have  assumed  that
it was sufficient to use one sequence of model photospheres based on a
representative  chemical composition
in analyzing low resolution spectra of cool dwarfs.
We must now admit that such an assumption is inappropriate, and we 
should consider abundance effects more carefully, 
especially of C \& O, in our future analysis of cool dwarfs.    
Also, we cannot use any solar composition for cool dwarfs
unless this substitution can be justified by a direct analysis
of the spectra of cool dwarfs.

It is true that a  detailed abundance analysis of brown dwarfs
is difficult especially with low resolution spectra, but well defined 
molecular bands, even  at low resolution, can be potential abundance 
indicators. We know already that CO$_2 $ is a fine indicator of 
C \& O abundances. Unfortunately, however, CO$_2$ is accessible only 
from space telescopes
and, moreover, spectroscopic observations in the near infrared are mostly
neglected by the recent space infrared missions.
From the view point of the study on cool dwarfs (and  other
cool stars), the importance of observing the near infrared
spectra (especially between 2.5 and 5.0\,$\mu$m) from space
cannot be emphasized too much.  

Although the spectra of brown dwarfs appear to be 
complicated, we  are now convinced that  the spectra of brown dwarfs can
 basically be understood on the basis of the LTE model photospheres, but 
only if the chemical composition is properly considered.
This is a reasonable result for such high density photospheres 
as of brown dwarfs in which frequent collisions easily maintain
 thermal equilibrium. 
Thus the chemical composition is the most important ingredient
in the interpretation and analysis of even  low resolution spectra.
Now, with better observed data for brown dwarfs including those
from space, analysis 
of the spectra and abundance determination can be done iteratively 
for brown dwarfs as for ordinary stars. 

Finally, we must remember that a major difficulty in  the analysis
of the spectra of brown dwarfs is that we have no model of comparable accuracy 
as for ordinary stars yet. For this reason, even the accurate numerical
method such as outlined in Section~\ref{sec:fitcmp}  cannot be 
infallible. In fact, we have no  model reproducing  all
the observable features correctly,  and the model found by the 
numerical method as well as by the eye-fitting method
may prove incorrect even if they are relatively satisfactory 
among the models  currently available. Within this limitation, we 
hope that  our main results on the differential effects of C \& O
abundances are relatively free of present brown dwarf model
uncertainties.

\acknowledgments
We thank an anonymous referee for critical reading of the text
and for invaluable suggestions regarding the method of analysis
of the spectra of brown dwarfs. We are grateful to Dr. Poshak Gandhi
for his careful checking of the manuscript and many suggestions to
improve the text.
This research is based on observations with \textit{AKARI},
a JAXA project with the participation of ESA.
We acknowledge JSPS/KAKENHI(C) No.22540260 (PI: I. Yamamura).

\end{document}